\documentclass[twocolumn,showpacs,amsmath,amssymb,superscriptaddress]{revtex4} 

\usepackage{graphicx}
\usepackage{dcolumn}
\usepackage{bm}

\newcommand{\bmu}{\mbox{\boldmath${\mu}$}}           
\newcommand{\bxi}{\mbox{\boldmath${\xi}$}}   

\begin{document}

\title{Propulsion with a Rotating Elastic Nanorod.}

\author{Manoel Manghi}
\affiliation{Laboratoire de Physique Th\'eorique, IRSAMC, Universit\'e Paul Sabatier, 31062 Toulouse, France}\author{Xaver Schlagberger}\author{Roland R. Netz}
\affiliation{Physik Department, Technical University Munich, 85748 Garching, Germany}

\date{\today}

\begin{abstract}
The dynamics of a rotating elastic filament is investigated using Stokesian simulations. 
The filament, straight and tilted with respect to its rotation axis for small driving torques, undergoes at a critical torque a strongly discontinuous shape bifurcation to a helical state. It induces a substantial forward propulsion whatever the sense of rotation: a nanomechanical force-rectification device is 
established.
\end{abstract}

\pacs{87.19.St, 47.15.Gf, 05.45.-a}

\maketitle

Propulsion of a micrometer sized object through a viscous fluid, where friction dominates and motion is invariably overdamped, calls for design strategies very different from the ordinary macroscopic world, where inertia plays an important role~\cite{purcell}. Biology has come up with a large number of concepts, ranging from active polymerization of gel networks~\cite{Joanny}, molecular motors moving on railway tracks formed by protein filaments~\cite{Prost}, to rotating or beating propellerlike appendices~\cite{lighthill}. Those propellers involve stiff polymers that are moved by molecular motors and rely on two different basic designs: i) Flagella of bacteria (prokaryotic cells) are helical stiff polymers set in motion at their base by a rotary motor. Hydrodynamic friction converts the rotational motion of the helix into thrust along the helix axis~\cite{lighthill,purcell}. ii) Cilia and flagella of sperms (eukaryotic organisms) are rodlike polymers that are anchored to a surface and beat back and forth driven by internal motors. They are used by small cells to swim but also by internal organs to pump liquids~\cite{lighthill}.

For a number of biomedical applications, e.g. for directed motion of artificial viruses through cells
or nano-devices through the bloodstream, as well as for mixing strategies in nanochips, it is desirable to develop similar synthetic propulsion mechanisms or to incorporate biological single-molecule motors into synthetic environments. Recent discoveries opened the route to the synthetic manufacture of rotary single-molecule motors driven by chemical~\cite{kelly} or optical~\cite{koumura,harada} energy. 
An ATPase can also be fixed to different substrates and used to rotate metallic~\cite{Soong} or organic nanorods~\cite{Noji}. This raises the question about the minimal design necessary to convert the rotational power of such nanoengines into directed thrust in a viscous environment. In this Letter, we show that an elastic rod that rotates around a point at a constrained azimuthal angle gives rise to a substantial forward thrust \emph{regardless of the sense of its rotation}. It thus acts as a rectification device and produces net thrust even when it is stochastically rotated back and forth. It makes usage of a helical polymer unnecessary, allowing a selection from the much wider class of straight stiff polymers. A few theoretical works have elucidated the coupling between hydrodynamics in a viscous medium and elasticity of soft polymers~\cite{wiggins,camalet,goldstein}. It was shown that finite stiffness of beating straight filaments breaks time-reversal symmetry and enables propulsion~\cite{wiggins,goldstein}. Conversely, reversing the rotation sense of bacterial flagella gives rise to pronounced shape polymorphic transformations~\cite{turner} and shows that elasto-hydrodynamics are important in biology. Likewise, hydrodynamic interactions between rotating flagella are involved in bundling and are important for switching between tumbling and running modes of \textit{E. coli} bacteria~\cite{ramia,Powers}.

We study an elastic rod whose base rotates on a cone driven by an external torque applied to its anchoring point. We use Stokes-simulation techniques where hydrodynamics are treated on the Rotne-Prager Green's function level including the full coupling between thermal, elastic and hydrodynamic forces. We obtain the nonlinear relation between driving torque and angular frequency in the presence of an external force load and estimate the power conversion efficiency. The filament exhibits a strongly discontinuous shape bifurcation at a finite torque value at which the rod jumps closer to the rotation axis.
\begin{figure*}[t]
\includegraphics[height=8.6cm]{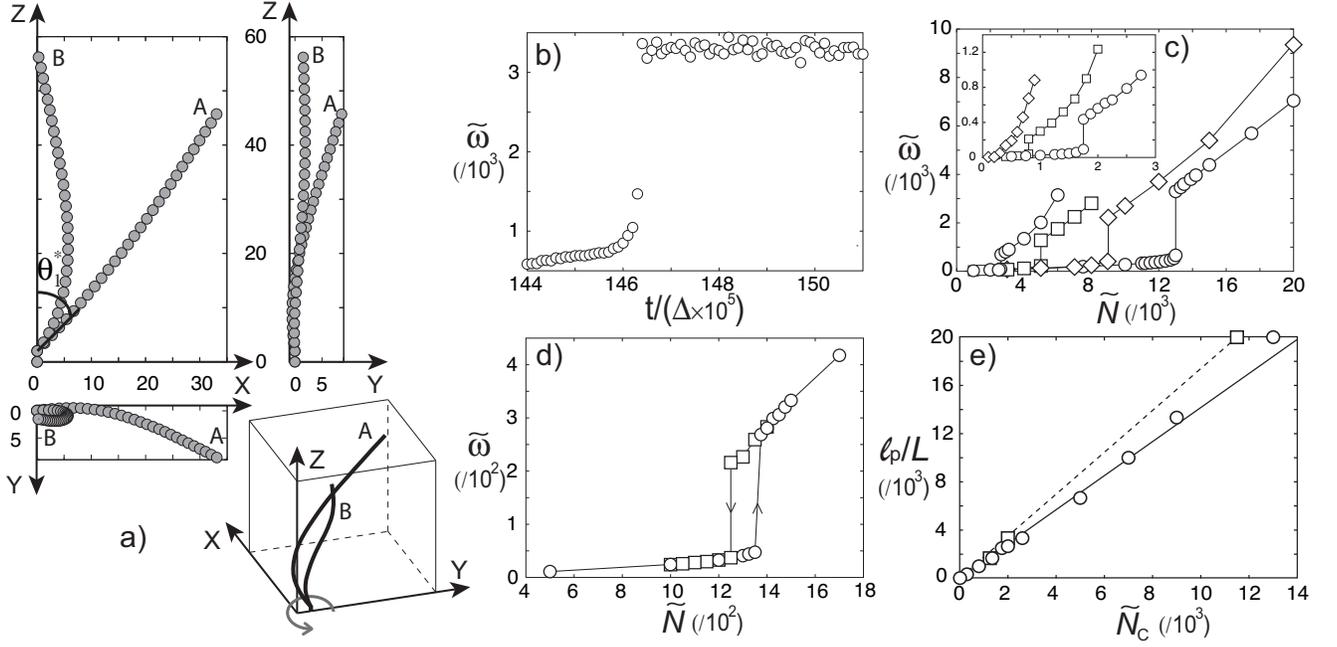} 
\caption{\label{omega}
a) Sketch in 3D and projected views of the rotating filament before (A, $\tilde{N}=600$) and after (B, $\tilde{N}=800$) the bifurcation ($M=30$ beads, $\ell_{\mathrm{p}}/L=10^3$). b) Angular velocity $\tilde{\omega}$ across the shape transition as a function of time ($\tilde{N}=1.33\times10^3$, $\ell_{\mathrm{p}}/L=2\times10^4$). c) Angular velocity as function of external torque, $\tilde{N}$, for various $\ell_{\mathrm{p}}/L$: 3333 ($\circ$), 6667 ($\square$), $1.33\times10^4$ ($\diamond$) and $2\times10^4$($\circ$) [inset: 333 ($\diamond$), 1000 ($\square$) and 2500 ($\circ$)]. d) Hysteresis cycle for $\ell_{\mathrm{p}}/L=1667$ for increasing  ($\circ$) and decreasing torque ($\square$) at constant rate $\delta \tilde{N}/(\delta t/\Delta)=1.25\times10^{-6}$. e) Persistence length vs. critical torque for increasing ($\circ$) and decreasing  torque ($\square$) both obeying a linear law according to Eq.~(\ref{scaling prediction}) (All data for stalled case, $V_z=0$).}
\vspace{-0.4cm}
\end{figure*}
The elastic polymer is modeled as a chain of $M+1$ connected spheres and the time evolution of the sphere positions $\mathbf{r}_i(t)$ is governed by the position-Langevin equation~\cite{ermack}
\begin{equation}
\dot{\mathbf{r}}_i(t)=-\sum^M_{j=0}\bmu_{ij}({\mathbf{r}_{ij}})\nabla_{\mathbf{r}_j}U({\mathbf{r}_k})+\mu_{i2}\mathbf{F}^{\mathrm{rot}}_2+ \bxi_i(t)
\label{langevin}
\end{equation}
where the potential $U({\mathbf{r}_k})=\sum^{M-1}_{i=0}[\frac{\gamma}{4a}(r_{ii+1}-2a)^2+\frac{\varepsilon}{2a}(1-\cos\theta_i)]$ is the discrete version of the extensible wormlike chain model, $a$ the sphere radius, $\mathbf{r}_{ij}=\mathbf{r}_j-\mathbf{r}_i$, and $\theta_i$ is the angle between neighboring bonds of sphere~$i$. The parameters $\gamma$ and $\varepsilon$ are the stretching and bending moduli, which are related by $\varepsilon/\gamma=a^2/4$ (isotropic rod). Hydrodynamic interactions between two spheres $i$ and $j$ are implemented via the Rotne-Prager mobility tensor 
\begin{equation}
\bmu_{ij}({\mathbf{r}_{ij}})= \frac{1}{8\pi\eta r_{ij}}\left[\mathbf{1}+\frac{\mathbf{r}_{ij}\mathbf{r}_{ij}}{r_{ij}r_{ij}}+\frac{2a^2}{r_{ij}^2}\left(\frac{\mathbf{1}}{3}-\frac{\mathbf{r}_{ij}\mathbf{r}_{ij}}{r_{ij}r_{ij}}\right)\right]
\end{equation}
which is an approximation valid for large sphere separations~\cite{ermack}. $\mathbf{1}$ denotes the unit $3\times3$ matrix and $\eta$ the viscosity of the aqueous solvent. For the self-mobility we choose the Stokes mobility of a sphere $\mu_{ii}=\mu_0=1/(6\pi\eta a)$. To mimic a rotary motor, we apply an external force $\mathbf{F}^{\mathrm{rot}}_2$ on monomer 2, which is related to the torque $\mathbf{N}$ by $\mathbf{F}^{\mathrm{rot}}_2=\mathbf{N}\times\mathbf{r}_{12}/r_{12}^2$. The Langevin random displacement $\bxi(t)$ models the action of the heat bath and obeys the fluctuation-dissipation relation
$\left\langle\bxi_i(t)\bxi_j(t')\right\rangle=2k_BT\bmu_{ij}\delta(t-t')$ numerically implemented by a Cholesky decomposition~\cite{ermack}. The persistence length is given by $\ell_{\mathrm{p}}=\varepsilon/(k_BT)$. Twist and torsional degrees are omitted since for most synthetic polymers, free rotation around the polymer backbone is possible. Two different force ensembles are investigated: the \textit{stalled} case, where the first two monomers forming the base are held fixed in space by applying virtual forces which exactly cancel all other elastic and hydrodynamic forces. In the \textit{moving} case, we let the polymer move along the $z$ axis and apply virtual forces on the first two monomers only laterally such that the polymer base moves along a vertical line. Hydrodynamic boundary effects are considered further below. 
A finite tilt angle at the base is imposed by a spontaneous curvature term in the elastic energy $(\varepsilon/2a)[1-\cos(\theta_1-\theta^*_1)]$; in this paper we show data for spontaneous curvature  $\theta^*_1=45^{\circ}$. For the numerical iterations, we discretize Eq.~(\ref{langevin}) with time step $\Delta$. By rescaling time, space and energies, we arrive at rescaled parameters $\tilde{\Delta}=\Delta k_BT\mu_0/a^2$, $\ell_{\mathrm{p}}/L=\varepsilon/(2aMk_BT)$ and $\tilde{N}=N/k_BT$. For sufficient numerical accuracy we choose time steps in the range $\tilde{\Delta} = 10^{-3}$--$10^{-8}$. Output values are calculated every $10^3$--$10^4$ steps, total simulation times are $10^8-10^9$ steps.
 
When a  torque is applied to the filament base, it  rotates and after some time exhibits a stationary shape. Since the friction is larger at the free end, the filament bends into a curved structure. The stationary angular velocity $\tilde{\omega}=\omega a^2/(k_BT\mu_0)$ (in degrees) is plotted in Fig.~\ref{omega}c as a function of the applied torque $\tilde{N}$ showing a nonlinear increase, which is caused by the finite bending rigidity. Indeed, for an infinitely stiff polymer, this relation is strictly linear. For a critical torque, $\tilde{N}_{\mathrm{c}}$, a shape bifurcation occurs and the angular frequency jumps dramatically. Figure~\ref{omega}a shows the stationary shape of the polymer before (A)  and after the bifurcation (B). The high-torque state is characterized by a smaller distance from the rotation axis. Figure~\ref{omega}b shows the actual time-dependent behavior of the rotational speed as the rod crosses the transition and settles in the stationary state. Pronounced hysteresis (which becomes slightly weaker with decreasing torque sweep rate) is observed when $\tilde{N}$ is varied across the critical region, as illustrated in Fig.~\ref{omega}d. Figure~\ref{omega}e shows the dependence of the critical torque on the persistence length of the rod. Since the torque change rate is finite, increasing ($\circ$) and decreasing torque ($\square$) gives rise to slightly different bifurcation values. For very small persistence lengths, conformational fluctuations suppress the transition. 
\begin{figure*}[t]
\includegraphics[height=4.2cm]{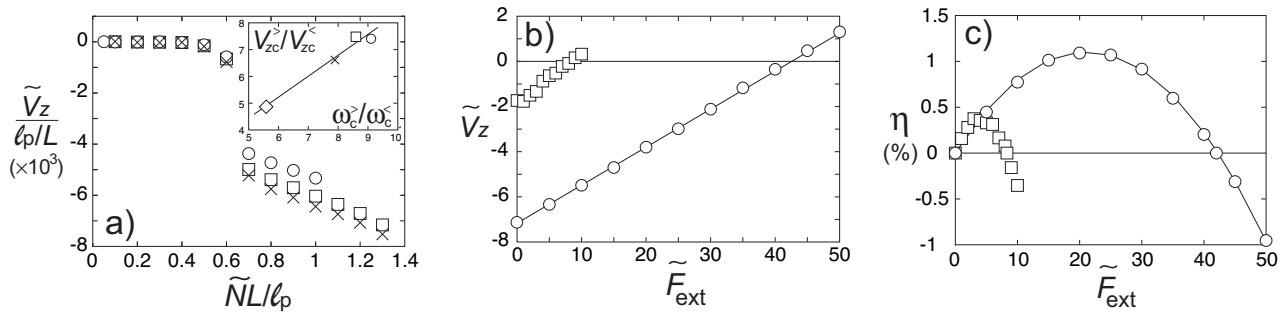} 
\caption{\label{propulsion} a) Propulsion velocity parallel to the rotation axis vs. $\tilde{N}L/\ell_{\mathrm{p}}$ in the $z$-\textit{moving} case for $\ell_{\mathrm{p}}/L=2500$ ($\times$), 5000 ($\square$) and $2\times10^4$ ($\circ$). The inset shows the variations of the ratio of velocities after and before the transition vs. the ratio of angular velocities ($\diamond$, $\ell_{\mathrm{p}}/L=1667$). b) $\tilde{V}_z$ vs. the external force $\tilde{F}_{\mathrm{ext}}$ applied on monomer 2 ($\ell_{\mathrm{p}}/L=1667$) before the bifurcation for $\tilde{N}=0.78 \ell_p/L$ ($\square$) and after 
$\tilde{N}=0.84 \ell_p/L$ ($\circ$). c) Efficiency, $\eta$, vs. $\tilde{F}_{\mathrm{ext}}$ (same parameters as b). The solid line is a  parabolic fit.}
\vspace{-0.4cm}
\end{figure*}
The shape bifurcation can be understood by simply balancing elastic and driving torques: the bending torque due to the filament deformation, projected along the vertical axis, reads $\sim \varepsilon\sin\theta_1/R$ where $R$ is the bending radius of the filament, which has to be balanced by the external torque $N$. The transition occurs when the bending radius reaches the length of the filament, $R \sim L$. This yields the critical torque 
\begin{equation}
\tilde{N}_{\mathrm{c}}\simeq\frac{\ell_{\mathrm{p}}}{L}\sin\theta_1
\label{scaling prediction}
\end{equation}
which is shown in Fig.~\ref{omega}e as a solid line and agrees very well with the numerical data.  
Neither hydrodynamical parameters nor temperature appear in Eq.~(\ref{scaling prediction});
the bifurcation is thus purely \textit{elastic} in origin and indeed can be obtained within free-draining or slender-body approximation (as will be published separately).
Using the rotational mobility $\mu_{\mathrm{rot}} \sim 1/(\eta L^3)$, the critical angular frequency reads 
$\omega_{\mathrm{c}} \sim \mu_{\mathrm{rot}} N_c  \sim \varepsilon /(\eta L^4)$. This threshold turns out to be much lower than for  the continuous twirling-whirling transition of a rotating rod with torsional elasticity, $\omega_{\mathrm{c}}^{\mathrm{tw}} \sim \varepsilon/(\eta a^2  L^2)$, which was obtained using linear analysis~\cite{goldstein}.

We define the translational and rotational mobilities of the filament as~\cite{purcell}
\begin{equation}
{\omega \choose V_z}  = \left( \begin{array}{cc} \mu_{\mathrm{rr}} & \mu_{\mathrm{rt}} \\ \mu_{\mathrm{tr}} & \mu_{\mathrm{tt}} \end{array}\right) {N \choose F_{\mathrm{ext}}}
\label{mobilities}
\end{equation}
where $V_z$ is the translational velocity downwards, $F_{\mathrm{ext}}$ the corresponding external force applied at the propeller base. In general, the mobilities $\mu$ depend on torque $N$ and force $F_{\mathrm{ext}}$ in a nonlinear fashion. For perfectly stiff propellers, the mobility matrix is constant and symmetrical, i.e. $\mu_{\mathrm{rt}}=\mu_{\mathrm{tr}}$~\cite{purcell}. However, in our case, due to flexibility, the propeller shape changes with external efforts and the matrix is asymmetric. The propulsion velocity along the rotation axis is plotted in Fig.~\ref{propulsion}a as a function of  $\tilde{N}L/\ell_{\mathrm{p}}$ for different persistence lengths for the $z$-\textit{moving} case ($F_{\mathrm{ext}}=0$). At the transition, a jump in the propulsion velocity is observed and $\tilde{V}_z$ is almost linear for $\tilde{N}>\tilde{N}_{\mathrm{c}}$, revealing that the shape remains almost fixed in this range of torque values. Inset of Fig.~\ref{propulsion}a shows the ratio of $V_z$ after and before the transition vs. the same ratio for $\omega$. The variation is roughly linear, meaning that the jump of $\tilde{V}_z$ is directly due to the jump in $\omega$. To test our propeller under load, we applied an external force, $\tilde{F}_{\mathrm{ext}}$, which we define to be positive when it pushes against its natural swimming direction. Figure~\ref{propulsion}b shows the variation of $\tilde{V}_z$ with $\tilde{F}_{\mathrm{ext}}$ at two different torques just below and above the bifurcation. The laws are almost linear in both cases, meaning that $\mu_{\mathrm{tt}}$ is almost independent of $F_{\mathrm{ext}}$. The efficiency of the power converter can be defined as the ratio of the propulsive power output and the rotary power input,
\begin{equation}
\eta=-\frac{F_{\mathrm{ext}} V_z}{N\omega}.
\label{efficiency}
\end{equation}
By inserting Eq.~(\ref{mobilities}) in Eq.~(\ref{efficiency}), we obtain $\eta(F_{\mathrm{ext}})=-(\mu_{\mathrm{tr}}NF_{\mathrm{ext}}+\mu_{\mathrm{tt}} F_{\mathrm{ext}}^2)/(\mu_{\mathrm{rr}}N^2+\mu_{\mathrm{rt}} N F_{\mathrm{ext}})$. We have checked that $\mu_{\mathrm{rt}}$ is negligibly small. Hence, the efficiency becomes parabolic as a function of the external force as shown in Fig.~\ref{propulsion}c. The highest efficiency is  obtained for $F_{\mathrm{ext}}=-\mu_{\mathrm{tr}}N/(2\mu_{\mathrm{tt}})=F_{\mathrm{stall}}/2$ and is only of the order of 1\%  after the transition and 3 times smaller before.

\begin{figure}[t]
\includegraphics[height=4.4cm]{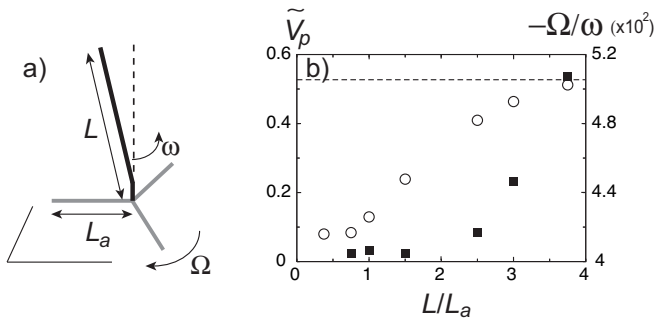} 
\caption{\label{device} a) Sketch of the self-propelling nanomachine whose base is made of 3 arms of length $L_a$. b) Propulsion velocity $\tilde{V}_p$ ($\circ$) and ratio of the base, $\Omega$, and filament, $\omega$, angular velocities ($\blacksquare$) against $L/L_a$ for $L=30 a$, $\ell_{\mathrm{p}}/L=33$, $\ell_{\mathrm{p}}^a/\ell_{\mathrm{p}} =10$ and $N=2000$. The dashed line is the asymptotic velocity reached by an isolated filament.}
\vspace{-0.4cm}
\end{figure}

Up to now we considered an isolated rotating filament to which external torque and forces were applied. In reality, the filament is attached to a base, and rotation of the filament is caused by some relative  torque generated between base and filament. As for any self-propelling object in viscous solvent, the total force and torque add up to zero~\cite{purcell}. If the filament rotates counterclockwise at frequency $\omega$, then the base rotates clockwise with an angular velocity $\Omega$. Within our Stokes simulation, we model the base by three elastic arms of length $L_a$ and persistence length $\ell_{\mathrm{p}}^a$ which are coplanar and connected to the filament at one point (see Fig.~\ref{device}a). Hydrodynamic screening and lubrication are accurately accounted for on scales larger than the monomer radius $a$. The torque acting on the filament is balanced by a counter torque on the base. The resulting complex helical trajectories of this nanomachine make calculation
of the mean propulsion velocity difficult due to thermally caused reorientation of the complete device. In the following, we therefore turn off the thermal noise (which plays a minor role for stiff polymers). Figure~\ref{device} shows the propulsion velocity $\tilde{V}_p$ and the ratio of rotational speeds of the base and the filament, $-\tilde{\Omega}/\tilde{\omega}$ vs. the ratio $L/L_a$ at fixed relative torque $\tilde{N}=2000$ ($L=30a$, $\ell_{\mathrm{p}}/L=33$, $\ell_{\mathrm{p}}^a/\ell_{\mathrm{p}} =10$). 
For base sizes comparable or larger than the filament length, $L/L_a < 1$, the propulsion velocity is highly reduced compared to the case of an isolated filament (dashed line), mostly due to hydrodynamic friction caused by the base. But for decreasing base size, the propulsion velocity approaches quite closely the velocity of an isolated filament, calculated according to the preceding model. Indeed, when it is very small, the counter-rotating base causes negligible hydrodynamic effects on the filament. At the same time, the base rotational speed $\Omega$ (rescaled by $\omega$, see Fig.~\ref{device}) increases when the base size decreases while remaining much smaller than the filament one. This is because, for such a torque much larger than the critical torque, the filament is close to the filament axis and thus gives little rotational resistance. It also means that only a small fraction of power input is invested into the base rotation.
In conclusion, we show that a simple straight elastic filament can induce propulsion when rotated at one end. The interplay between elastic deformations and hydrodynamic interactions results in a substantial directed thrust due to a subcritical dynamic bifurcation for control parameters $\tilde{N}/\tilde{N}_{\mathrm{c}}>1$. The mobility $\mu_{\mathrm{tr}}$ changes its sign when the torque is reversed, which implies a forward thrust whatever the sense of rotation. This propulsion occurs 
even in the presence of an explicit base, being larger for small bases. 
This work provides a clue for the synthetic manufacture of biomimetic micropropeller, by using simple semiflexible polymers instead of rigid helices. This bifurcation could be experimentally investigated using a macroscopic scale model similar to the one developed by Powers group~\cite{kim}. With their experimental setup, the bifurcation would occur for torques on the order of 0.05~Nm and angular velocities larger than 0.01~Hz, which are accessible.

With regard to bacteria propulsion, flagella are helical and their physics is therefore much more involved. However, flagellar motors generate sufficient torques for the nonlinear elastic phenomena discussed here: they are powered by a proton-motive force which yields torques $\sim 10^3\,\,k_BT$~\cite{berry,turner}. Moreover, the flagellum length is $L\simeq 10\;\mu$m, leading to $\ell_{\mathrm{p}}/L\simeq 10^3--10^4$ and $\tilde{N}L/\ell_{\mathrm{p}}\simeq 0.1--1$. Hence, the shape bifurcation discussed in this model is probably biologically relevant and might be directly observed with straight biopolymers attached to flagellar motors. For a bacterial flagellum at physiological conditions, the diameter is roughly $a\simeq 20$~nm and the stall forces are roughly 10~pN at the bifurcation. Critical angular and propulsion velocities (with no external force) then follow from our results as $\omega_{\mathrm{c}}\simeq 2\times10^5--10^6$ rad/s and $V_{z\mathrm{c}}\simeq 6--40$ mm/s. 
These values are large but if we actually consider a thicker bundle formed by approximately 7 flagella~\cite{turner}, we come up with smaller velocities which 
are comparable to the ones observed for \textit{E. coli} ($\omega\simeq10^4$~rad/s and $V_z\simeq 30\;\mu$m/s).

Financial support of the German Science Foundation (DFG, SPP1164) is acknowledged.


\begin{thebibliography}{99}
\bibitem{purcell} E.M. Purcell, PNAS, \textbf{94}, 11307 (1997); Am. J. Phys., \textbf{45}, 3 (1977).

\bibitem{Joanny} K. Kruse, J. F. Joanny, F. J\"ulicher and J. Prost, Phys. Rev. Lett., \textbf{92}, 078101 (2004).

\bibitem{Prost} F. J\"ulicher, A. Ajdari and J. Prost, Rev. Mod. Phys., \textbf{69}, 1269 (1997).

\bibitem{lighthill} J. Lighthill, SIAM Rev., \textbf{18}, 161 (1976).

\bibitem{kelly} T.R. Kelly, H. De Silva, R.A. Silva , Nature (London), 
\textbf{401}, 150 (1999).

\bibitem{koumura} N. Koumura et al., Nature (London), \textbf{401}, 152 (1999).

\bibitem{harada} T. Harada and K. Yoshikawa, Appl. Phys. Lett., \textbf{81}, 4850 (2002).

\bibitem{Soong} R.K. Soong et al., Science {\bf 290}, 1555 (2000).

\bibitem{Noji} H. Noji, R. Yasuda, M. Yoshida, K. Kinosita, Nature (London) {\bf 386}, 299 (1997).

\bibitem{wiggins} C.H. Wiggins and R.E. Goldstein, Phys. Rev. Lett., \textbf{80}, 3879 (1998).

\bibitem{camalet} S. Camalet, F. J\"ulicher, and J. Prost, Phys. Rev. Lett., \textbf{82}, 1590 (1999).

\bibitem{goldstein} C.W. Wolgemuth, T.R. Powers and R.E. Goldstein, Phys. Rev. Lett., \textbf{84}, 1623 (2000).

\bibitem{turner} L. Turner, W.S.  Ryu and H.C. Berg, J. Bacteriol., \textbf{182}, 2793 (2000).

\bibitem{ramia} M. Ramia, D.L. Tullock and N. Phan-Thien, Biophys. J., \textbf{65}, 755 (1993).

\bibitem{Powers} M. Kim and T.R. Powers, Phys. Rev. E {\bf 69}, 061910 (2004).

\bibitem{ermack} D.L. Ermack and J.D. McCammon, J. Chem. Phys., \textbf{69}, 1352 (1978).

\bibitem{kim} M. Kim et al., PNAS {\bf 100}, 15481 (2003).

\bibitem{berry} R.M. Berry, in \textit{Forces, growth and form in soft condensed matter: at the interface between physics and biology}, A. T. Skjeltorp and A. V. Belushkin eds. (Kluwer Acad. Publ., Dordrecht, 2004).
\end{thebibliography}
\end{document}